\documentclass[
aip, 
jmp,
amsmath, amssymb, 
reprint,
]{revtex4-1}

\usepackage{graphicx}
\usepackage{makeidx}
\usepackage{multirow}
\usepackage{amssymb}
\usepackage{amsfonts}
\usepackage{amsmath}
\usepackage{amsthm}
\usepackage{natbib}

\usepackage{graphicx}
\usepackage{dcolumn}
\usepackage{bm}

\usepackage[utf8]{inputenc}
\usepackage[T1]{fontenc}
\usepackage{mathptmx}
\usepackage{amsfonts}
\usepackage{bbm}
\usepackage[bottom]{footmisc}

\begin{document}

\preprint{AIP/123-QED}

\title{On reconstructing parts of quantum theory from two related maximal conceptual variables.}

\author{Inge S. Helland}
 \email{ ingeh@math.uio.no}
\affiliation{Department of Mathematics, University of Oslo \\ P.O.Box 1053 Blindern, N-0316 Oslo, Norway
\\ Orcid: 0000-0002-7136-873X}


\begin{abstract}

In the book [4] the general problem of reconstructing the Hilbert space formulation in quantum theory is discussed from the point of view of what I called conceptual variables, any variables defined by a person or by a group of persons. These variables may be inaccessible, i.e., impossible to assign numerical value to by experiments or by measurements, or accessible. One basic assumption in [4] and here is that  group actions $g\in G$ are defined on a space where some maximally accessible variable varies, and then accessible functions of these maximal variables are introduced. By using group representation theory the basic Hilbert space formalism is restored under the assumption that the observator or the set of observators has two related maximally accessible variables in his (their) mind(s). The notion of relationship is precisely defined here. Symmetric (self-adjoint) operators are connected to each variable, and in the discrete case the possible values of the variables are given by the eigenvalues of the operators. In this paper the main results from  [4] are made more precise and more general. It turns out that the conditions of the main theorem there can be weakened in two essential ways: 1) No measurements need to be assumed, so the result is also applicable to general decision situations; 2) States can have arbitrary phase factors. Some consequences of this approach towards quantum theory are also  discussed here.

\end{abstract}

\maketitle

\underline{Keywords:} Accessible conceptual variables; group actions; group representations; Hilbert space formulation; maximal conceptual variables; quantum theory,

\section{Introduction}

The foundation of and the interpretation of quantum mechanics have been discussed since the theory was initially developed more than 100 years ago. The discussion has been particularly intense in the last decade.

As is well known, the theory was originally two theories, the matrix theory of Heisenberg and the wave theory of Schr\"{o}dinger. These were shown to be mathematically equivalent, and von Neumann gave a well known and highly respected foundation in his famous book. This mathematical foundation has been reproduced in dozens of textbooks.

Take one of those textbooks, Ballentine [1]. According to this book and similar books, it is all based upon a few postulates. The basic notions are unit vectors describing states and selfadjoint operators describing physical variables. The postulates take as a basis a fixed Hilbert space describing the physical situation. Explicitely, two of Ballentine's postulates are
\bigskip

\textbf{Postulate 1a.} \textit{To each dynamic variable there is a Hermitian operator whose eigenvalues are possible values of the dynamic variable.}
\bigskip

\textbf{Postulate 2a.} \textit{To each state there corresponds a unique state operator, which must be Hermitian, nonnegative, and of unit trace.}
\bigskip

More concise mathematical formulations are given in Takhtajan [2] and Hall [3], where also state spaces corresponding to continuous variables are treated. In the latter books \emph{self-adjoint} is mostly used instead of \textit{Hermitian}. In [4] it is also shown that \textit{symmetric} operators under wide conditions are self-adjoint. In any case, these terms are defined in the books.

Below I will address a variant of these sets of postulates, first what I will call `Postulate' 1; inverted commas are used because I will show below that this follows from much simpler conditions. These conditions and the `postulate' take as a point of departure what I call conceptual variables, that are closely related to Ballentine's dynamical variables. The difference is explained below. This `postulate' together with the Born formula imply the other postulate listed above.
\bigskip

\textbf{`Postulate' 1.} \textit{To each conceptual variable there exists a unique symmetric operator. In the discrete case the set of eigenvalues equals the set of possible values of the conceptual variable. If and only if the variable is what will be defined below as maximally accessible, all the eigenspaces are one-dimensional. A pure state can be defined by a question involving a maximally accessible variable together with a sharp answer to this question. To each pure state there corresponds a unique one-dimensional projection operator, equivalently, a unit vector defined up to a phase factor.}
\bigskip

The purpose of this paper is to show how  `Postulate' 1 can be derived from simple assumptions on the conceptual variables. (Theorem 1 below is a part of this; the rest is essentially proved in [4].) This is called the Hilbert space theorem. An open question is to what extent the pure state definition given above covers everything that is considered as pure states in traditional quantum machanics. It does so in many important examples, however.

Also arguments behind the Born formula are given in [4]. 

The book [4] was explicit about what was called epistemic processes, processes to achieve knowledge through experiments or through measurements, and the Hilbert space theorem was there limited to that setting. But it was indicated that quantum theory also had wider applications; in particular, quantum decision theory was described.

In a first article related to the present paper, published early in November on arXiv, I indicated how parts of the Hilbert space theorem could be generalized. So far, I have received few reactions on the book and on the paper, but one reaction was a statement to the effect that the Hilbert space theorem could not be seen as a mathematical theorem. 

The additional purpose of the present paper is then twofold: 1) To give a better description of the relevant background and the notions and assumptions behind the theorem, 2) To show that it really is a mathematical theorem. The resulting discussion is also considerably extended here compared to the arXiv paper.

The reconstruction of the Hilbert space formulation proposed here, can not be compared directly to other recent reconstructions referred to in [4], but can be seen in conjunction with these. As far as I know, none of the other relevant papers use group representation to define the basic Hilbert space.

The theory presented here is almost complete. An open problem is to prove that different conceptual variables always give different operators.

In Sections II  and III of this paper the background is sketched, and in Section IV the basic definitions are given. For the plan of the rest of the paper, see Section IV.

\section{ A simple sketch of a process in empirical science.}

A scientist may be interested in asking many different questions to Nature. Some of these may be of the form: `What will the quantity $\theta$ turn out to be if I  measure it?' In connection to such questions, I will call $\theta$ a conceptual variable. In a concrete context there may be many such variables, in general $\{\theta^a ; a\in\mathcal{A}\}$ for some index set $\mathcal{A}$. 

Assume that he concentrates on one such $\theta$. His first task is to set up an experiment or a measurement. This will result in some data that hopefully give some information on $\theta$. If $\theta$ is a scalar, this information may lead to an interval which we claim that $\theta$ is contained in, in statistical language a confidence interval or a Bayesian credibility interval. If $\theta$ is discrete and we have good data, these intervals may collapse into a single point, and we end up with conclusions of the form $'t=u'$, where $u$ is a possible value. In my view, this may be taken as a description of a measurement related to elementary quantum mechanics.

\section{Further background and assumptions.}

The basic notion of [4] and the present paper is that the essence of quantum theory in some given context can be described in terms of conceptual variables in the mind of an observer or in the joint minds of a group of communicating observers. In the epistemic process situation these are called epistemic conceptual variables, or simply e-variables. I will start with a simple example.

Let us assume that we are given some object A, and ask 'What is the weight of object A?'. Then $\mu=$'weight of A' is an e-variable. Of course it exists as a variable in the real world, but in my opinion it must also during measurement be said to exist in the mind of the relevant observer. Every conveivable model of the situation is created by this mind. We can use a scale to obtain a very accurate estimate of $\mu$. Or we can use several independent measurements, and use the mean of those as a more accurate estimate. In the latter case it is common to introduce a statistical model where $\mu$ is a parameter of that model. But in my view the e-variable concept is a more fundamental notion. The variable $\mu$ exists before any statistical model is introduced. Most people will agree that $\mu$ exists in some sense. In this example the e-variable has some ontic basis, but my claim is that this need not always be the case in all epistemic processes. Even in this case the existence of $\mu$ as a real number may be discussed. For instance, the question 'Is $\mu$ rational or irrational?' is rather meaningless.

This example is carried over to a general measurement situation. The crucial assumption is that \emph{e-variables have a double existence: One existence which in some sense may be connected to the real world, and one existence in the mind of a relevant observer or a group of observers.}

This assumption can in some way be said to be inspired by statistical theory: Basic statistical theory relies on statistical models of data, given some parameters. These models are formulated in attempts to descibe the real situation at hand as well as possible. If necessary, they can be tested against data. The models are always formulated in the mind of some researcher or in the joint minds of a group of researchers.  In particular, the concept of \emph{parameter} in statistics, a concept that was introduced in an essential way by Ronald Fisher, has a basis which is related to this mind/ these minds. Unfortunately, the word `parameter' also has other meanings in the physical literature; therefore it is avoided here, and it is also avoided in physical contexts in [4].

A fundamental problem in quantum theory is the question of whether or not it is possible to derive the Hilbert space formulation from other, more intuitive axioms and notions. In [4] this problem was approached by using basic group theory together with the notion of accessible and inaccessible conceptual variables. At the outset, a conceptual variable in some given setting is any variable defined by a person or a group of persons. In a typical quantum context the setting may be one where an observer seeks some knowledge about a physical system, either through experiments or through some measurement. In [4] and [5] this is called an epistemic process. The variable may then simply be a physical variable as defined by the observer. Essentially different situations are quantum models of cognition and decision (see [6]), where the variables may have a psychological origin.

One limitation of the derivation in [4] is that, basically, an epistemic prosess is assumed there, so that general applications to cognitions and decisions are not covered. Here we will try to improve this, in particular, we will make no assumption that there exists some measurement of the basic accessible conceptual variable. This has important consequences: The whole theory may be generalized to situations determined by cognition and decision; see [6] and [7]. 

Another limitation was that the proof of the basic Theorem 4.3 in [4] was based on the following simple version of quantum theory: Each (maximally accessible; see below) state is given by a single unit vector in a Hilbert space. In reality the state is determined by a group of unit vectors \emph{with arbitrary phase factors}. This will be tried taken into account here.

\section{On accessibility and beyond}

A conceptual variable is called \emph{accessible} if it is possible for an agent to assign a numerical value to it by a suitable empirical process; it is called \emph{inaccessible} if this is not possible. In a physical setting an inaccessible  variable might be a vector (position, momentum) or the total spin vector of a particle. A psychological case may be a decision situation where the decision variable is so comprehensive that the agent is unable to make a decision. As in [4], a given inaccessible variable is called $\phi$, and is thought to vary in some space $\Omega_\phi$. Group actions $k\in K$ may be supposed to be defined on $\Omega_\phi$.

Accessible variables which can be seen as functions on $\Omega_\phi$:  $\theta=\theta(\phi)$ were studied in [4] and [8], and $\theta$ was assumed to vary on a space $\Omega_\theta$. In a physical situation, $\theta$ can, for instance be the spin component in some given direction; in a decision situation, it can be a simpler decision variable that enables the agent to make a partial decision. If $\phi=(\theta_1,\theta_2 )$, where $\theta_1$ and $\theta_2$ are accessible, but $\phi$ is inaccessible, we say that $\theta_1$ and  $\theta_2$ are \emph{complementary}.

One further notion is of interest:
\bigskip

\textbf{Definition 1.} \textit{A partial ordering is defined first on all conceptual variables and then on the accessible conceptual variables as follows: $\theta$ is said to be less than or equal to $\lambda$ if one can find a function $f$ such that $\theta(\phi) = f(\lambda(\phi))$. }
\bigskip

Note that if $\theta$ is accessible, and $\xi=h(\theta)$ for some function $h$, then $\xi$ is accessible. Hence this is really a partial ordering on the accessible variables. By Zorn's Lemma there then exist \emph{maximally accessible conceptual variables}, and these will be important in what follows.

In certain connections also the following is important:
\bigskip

\textbf{Definition 2.} \textit{The variable $\theta$ is called permissible if the following holds: $\theta(\phi_1)=\theta(\phi_2)$ implies $\theta(k\phi_1)=\theta(k\phi_2)$ for all $k\in K$.}
\bigskip

If $\theta$ is permissible, one can define group actions $g\in G$ on $\Omega_\theta$ such that

\begin{equation}
(g\theta)(\phi)=\theta (k\phi);\ k\in K.
\label{G}
\end{equation}

The mapping from $K$ to $G$ defined by (\ref{G}) is a group (action) homomorphism. The left and right invariant measures under $K$ on $\Omega_\phi$ and correspondingly under $G$ on $\Omega_\theta$ may be defined under weak conditions.

Note that Definition 1 and Definition 2 are precise mathematical definitions, and can be taken as a basis for a precise mathematical theory. The only assumed axiom is that connected to the introduction of the accessibility-notion.

The point of departure in [4] is group representation theory and the theory of coherent states. Let $g\in G$ be group actions on $\Omega_\theta$, not necessarily induced by $K$ as above. Assume that $G$ is transitive, has a trivial isotropy group, and that $\theta$ is a maximally accessible variable. Introduce an irreducible unitary representation of the abstract group behind $G$: $\{U(g)\}$ acting on a Hilbert space $\mathcal{H}$, which for instance may be a closed subspace of some $L^2$-space. Fix some vector $|\theta_0 \rangle \in\mathcal{H}$ and  define the coherent states $|\theta\rangle=U(g)|\theta_0\rangle$ whenever $\theta=g\theta_0$ and $\theta_0 \in \Omega_\theta$ is fixed. Since there by assumptions is a one-to-one correspondence between $g$ and $\theta=g\theta_0$, this notation is in order: $U(\cdot)$ can be seen as a function from $\Omega_\theta$ to $\mathcal{H}$. From this, appealing to Schur's Lemma, a resolution of the identity

\begin{equation}
I=\int |\theta\rangle\langle\theta |d\rho (\theta)
\label{identity}
\end{equation}
is proved in [4]. Here $\rho$ is a left-invariant measure on $\Omega_\theta$, which under weak assumptions [1] will exist. Using this, an operator for the variable $\theta$ is defined as

\begin{equation}
A=\int\theta  |\theta\rangle\langle\theta |d\rho (\theta),
\label{A}
\end{equation}
assuming that this integral converges. 
If several accessible variables are to be discussed, it is useful to write the resolution of the identity in equivalent form

\begin{equation}
I=\int |\theta(\phi)\rangle\langle\theta(\phi) |d\mu (\phi),
\label{identity1}
\end{equation}
where $\mu$ is a left-invariant measure on $\Omega_\phi$ if such a measure exists. The operator for $\theta$ can then be written

\begin{equation}
A=\int\theta(\phi)  |\theta(\phi)\rangle\langle\theta(\phi) |d\mu(\phi),
\label{A1}
\end{equation}

If $G$ is not transitive at the outset, a model reduction is first made to an orbit of $G$ on $\Omega_\theta$. This was motivated in [4] by a similar procedure from statistics. 

Note that $A$ here is determined by the accessible variable $\theta$. Ideally, i want this functional dependence to be a one-to-one correspondence. Then there is a unique operator corresponding to every accessible conceptual variable, and we are in a setting usually assumed by quantum theory. To prove this uniqueness in general, is considered an open problem.

In order that the above development shall lead to a useful theory, several assumptions must be made: 1) $G$ is transitive on $\Omega_\theta$, and $\theta$ is maximally accessible; 2) The representation involved is irreducible; 3) There can be established a one-to-one correspondence between the coherent states $|\theta\rangle =U(g)|\theta_0\rangle$ on the one hand and the group elements $g$ and hence the values $\theta=g\theta_0$ of the conceptual variable on the other hand. In Section V below, the assumption 2) will be dropped, but instead it will be assumed that \emph{two} related maximally accessible variables are present.

Some of the main results of [4] are the following: If $A$ has a discrete spectrum, the possible values of $\theta$ are the eigenvalues of $A$, and $G$ is a subgroup of the group of permutations on these values. Furthermore, $\theta$ is maximally accessible if and only if each eigenspace of $A$ is one-dimensional.

Different (complementary) $\theta$ give different resolutions of the identity. If each such $\theta$ is discrete and  maximally accessible, each eigenvector can be interpreted as connected to the question `What is the value of $\theta$?' together with a sharp answer to it. In my opinion this can be used to give a simple interpretation of at least many quantum states. I will also consider operators for non-maximally accessible $\theta$; then the eigenspaces of $A$ have a similar interpretation. The generality of such interpretations for \emph{all} state vectors was discussed and also taken up as an open question in [9].

The purpose of this paper is to discuss some of this theory in more detail. The remainder of this paper is organized as follows: In Section V the assumptions behind the theory in [4] are briefly reviewed a little more precisely. Section VI gives the main theorem of [4] in a weaker and more precise form, and this theorem is proved in Section VII. In Section VIII the properties of the constructed operators are presented, and section IX  treats the spin case. Section X discusses the epistemic interpretation, and Section XI provides a final discussion.

The result of Section VI may possibly imply a completely new approach to quantum theory. This result can be seen as fundamental for the ideas promoted in [4] and also for the discussion of the Bell experiment in [10]. The development given here could ideally replace parts of Chapter 4 in this book, but that Chapter is fairly satisfactory as long as we limit ourselves to epistemic processes.

\section{Technical assumptions and basic theory}

$\Omega_\phi$ and $\Omega_\theta$ are assumed to be equipped with topologies, and all functions are assumed to be Borel measurable. The group actions are assumed to be locally compact.  As is common, it is assumed that the group operations on the underlying abstract groups: $(g_1,g_2)\mapsto g_1g_2$, $(g_1,g_2)\mapsto g_2 g_1$ and $g\mapsto g^{-1}$ are continuous. Furthermore, it is assumed that the action $(g,\theta)\mapsto g\theta$ is continuous. An additional assumption, which ensures the existence of invariant measures on $\Omega_\theta$, is the following: Every inverse image of compact sets under the mapping $(g,\theta)\mapsto (g\theta,\theta)$ should be compact. This ensures that a left-invariant measure $\rho$ on $\Omega_\theta$ exists.

Basic group representation theory was reviewed in [4]. It was assumed that the groups behind $K$ and $G$ have representations that give square-integrable coherent state systems (see [11]). It was assumed further that $G$ is transitive on $\Omega_\theta$ and that the isotropy group is trivial. Then, we can fix $\theta_0 \in \Omega_\theta$, and represent every $\theta \in \Omega_\theta$ by $\theta=g\theta_0$, and from this there is a one-to-one correspondence between $\theta$ and $g$. Let $U$ be a relevant irreducible unitary representation of the abstract group behind $G$ on a Hilbert space $\mathcal{H}$. Fix a vector in $\mathcal{H}$ and call it $|\theta_0\rangle$. The coherent states are then given by $U(g)|\theta_0\rangle$, and I define $|\theta\rangle = U(g)|\theta_0\rangle$ when $\theta=g\theta_0$. From this, the resolution of the identity (\ref{identity}) is proved in [4], and it is natural to define the operator $A$ corresponding to the variable $\theta$ by (\ref{A}). 

In general the operator $A$ of (\ref{A}) is unbounded, and is defined on the space of $|u\rangle\in\mathcal{H}$ for which $\langle u|A|u\rangle$ converges.

By the spectral theorem, $A$ has a spectral decomposition. As is proved in [4], if $A$ has a discrete spectrum, the possible values of $\theta$ are the eigenvalues of the operator.

To arrive at quantum theory from this basis, we need a rule to calculate probabilities, the Born rule. A recent review of various derivations of the Born rule is given in [12]. In [4] the rule is derived from three assumptions: 1) a focused likelihood principle, that follows from the ordinary likelihood principle of statistical inference; 2) an assumption of rationality as formulated by the Dutch book principle. 3) The variables involved are maximally accessible. In 2) it is not necessarily assumed that the actor(s) involved is (are) perfectly rational; it is enough that they have ideals which can be modelled in terms of an abstract perfectly rational being.

\section{The main theorem}

Theorem 4.3 in [4] was fundamental for that book. Here we want to prove a more precise and in some respects stronger  version of that result:
\bigskip

\textbf{Theorem 1.}
\textit{Consider a situation where there are two maximally accessible conceptual variables $\theta$ and $\xi$ in the mind of an actor or in the minds of a communicating group of actors. Make the following assumptions:}

\textit{(i) On one of these variables, $\theta$, there can be defined transitive group actions  $G$ with a trivial isotropy group and with a left-invariant measure $\rho$ on the space $\Omega_\theta$.}

\textit{(ii) There exists a unitary irreducible representation $U(\cdot)$ of the group behind the group actions $G$ defined on $\theta$ such that the coherent states $U(g)|\theta_0\rangle$ are in one-to-one correspondence with the values of $g$ and hence with the values of $\theta$.}

\textit{(iii) The two maximally accessible variables $\theta$ and $\xi$ can both be seen as functions of an underlying inaccessible variable $\phi\in\Omega_\phi$. There is a transformation $k$ acting on $\Omega_\phi$ such that $\xi(\phi)=\theta(k\phi)$.}

\textit{Then there exists a Hilbert space $\mathcal{H}$ connected to the situation, and to every accessible conceptual variable there can be associated a symmetric operator on $\mathcal{H}$.}
\bigskip

It is important that the theorem is no longer limited to an epistemic process; the variables $\theta$ and $\xi$ can also be general decision variables, variables connected to sets of prospects as defined in [7].

The condition (iii) seems to be crucial. Two variables $\theta$ an $\xi$ satisfying this condition are said to be  \emph{related}. If no such underlying variable $\phi$ can be found, the two variables are said to be \emph{essentially different}. 

The Hilbert space $\mathcal{H}$ is of course the Hilbert space associated with the representation (ii). The operators for maximally accessible variables are given for instance  by (\ref{A}). For accessible variables which are not maximal, we rely on the spectral theorem. For any accessible variable $\eta$ we can by the partial ordering find a function $f$ and a maximally accessible variable $\theta$ such that $\eta=f(\theta)$. Let the operator $A$, say from (\ref{A}), have a spectral representation
\begin{equation}
A=\int_{\sigma(A)}\theta dE(\theta),
\label{spectr}
\end{equation}
where $\sigma(A)$ is the spectrum of $A$, and $E$ is a projection-valued measure (see [4]). Then we can define
\begin{equation}
A^{\eta} = \int_{\sigma(A)} f(\theta) dE(\theta).
\label{etahat}
\end{equation}

The condition (ii) is rather technical, but essential for the proof of Theorem 1. The following result may be helpful.
\bigskip

\textbf{Proposition 1.} \textit{Let $\rho$ be the left-invariant measure on $\Omega_\theta$, and assume that}
\begin{equation}
\int |f(g\theta)-f(\theta)|d\rho(\theta)\ne 0
\label{v1}
\end{equation}
\textit{for every $f\in L^2(\Omega_\theta ,\rho)$ and every $g\in G$ different from the unit element.}

\textit{Then condition (ii) holds for the case where $\mathcal{H}$ is a closed subspace of $L^2(\Omega_\theta ,\rho)$ and $U$ is any irreducible subrepresentation of the regular representation given by $U(g) f(\theta)=f(g^{-1}\theta)$.}
\bigskip

\textit{Proof.}  We need to prove that $U(g_1) f (\theta) \ne U(g_2) f (\theta)$ whenever $g_1 \ne g_2$. In the $L^2$-space language and in terms of the regular representation this means that
\begin{equation}
\int |f(g_2^{-1}\theta)-f(g_1^{-1}\theta)|d\rho(\theta)\ne 0
\label{v2}
\end{equation}
But this follows from (\ref{v1}), since by left invariance $d\rho(\theta)=d\rho(g_1^{-1}\theta)$.
 
\qed

A consequence of this is that one in the discrete case may drop the assumption (ii) when $U(\cdot)$ is irreducible.

\section{Proof of the main theorem.}

\subsection{Basic construction}

In this subsection I will take as a point of departure Chapter 2 in Perelomov [11], which discusses coherent states for arbitrary Lie groups. Let $G$ be a transitive group action on the space $\Omega_\theta$ associated with some conceptual variable $\theta$ and $T(g)$ the unitary representation of its abstract group, acting on the Hilbert space $\mathcal{H}$. I will assume that $G$ has a trivial isotropy group, so that the elements $g$ of $G$ are in one-to-one correspondence with the values of $\theta$.

As in [11] (and in [4]) take a fixed vector $|\theta_0\rangle$ in $\mathcal{H}$, and consider the set $\{|\theta\rangle\}$, where $|\theta\rangle = T(g)|\theta_0\rangle$ with $g$ corresponding to the value $\theta$. It is not difficult to see that two vectors $|\theta^1\rangle$ and $|\theta^2\rangle$ correspond to the same state, i.e., differ by a phase factor ($|\theta^1\rangle =\mathrm{exp}(i\alpha)|\theta^2\rangle$, $|\mathrm{exp}(i\alpha)| =1$), only if $T(g^{2, -1}g^1)|\theta_0\rangle=\mathrm{exp}(i\alpha)|\theta_0\rangle$, where $g^1$ corresponds to $\theta^1$ and $g^2$ corresponds to $\theta^2$. Suppose $E=\{e\}$ is a subgroup of the group $G$, such that its elements have the property

\begin{equation}
T(e)|\theta_0\rangle =\mathrm{exp}[i\alpha(e)]|\theta_0\rangle .
\label{u1}
\end{equation}

When the subgroup $E$ is maximal, it will be called the isotropy subgroup for the state $|\theta_0\rangle$. More precisely, it is the isotropy subgroup of the group $T(G)$ corresponding to this state.

The construction shows that the vectors $|\theta\rangle$ corresponding to a value $\theta$ and thus to an element $g\in G$, for all the group elements $g$ belonging to a left coset class of $G$ with respect to the subgroup $E$, differ only in a phase factor and so determine the same state. Choosing a representative $g(x)$ in any equivalence class $x$, one gets a set of states $\{|\theta_{g(x)}\rangle\}$, where $x\in X=G/E$. Again, using the correspondence between $g$ and $\theta$, I will write these states as $\{|\theta(x)\rangle\}$, or in a more concise form $\{|x\rangle\}$, $|x\rangle\in \mathcal{H}$. 
\bigskip

\textbf{Definition 3.} \textit{The system of states $\{|\theta\rangle =T(g)|\theta_0\rangle\}$, where $g$ corresponds to $\theta$ as above, is called the coherent-state system $\{T,|\theta_0\rangle\}$. Let $E$ be the isotropy subgroup for the state $|\theta_0\rangle$. Then the coherent state $|\theta(g)\rangle$ is determined by a point $x=x(g)$ in the coset space $G/E$ corresponding to $g$ and to $|\theta(g)\rangle$ is defined by $|\theta(g)\rangle=\mathrm{exp}(i\alpha)|x\rangle$, $|\theta_0\rangle=|0\rangle$.}
\bigskip

\textbf{Remark.} The states corresponding to the vector $|x\rangle$ may also be considered as a one-dimensional subspace in $\mathcal{H}$, or as a projector $P_x =|x\rangle\langle x|$, $\mathrm{dim}\ P_x=1$, in $\mathcal{H}$. Thus the system of coherent states, as defined above, determines a set of one-dimensional subspaces in $\mathcal{H}$, parametrized by points of the homogeneous space $X=G/E$.
\bigskip

The general properties og the coherent state system was studies more closely in [11], and I refer to that discussion here. The important consequence for me is the following: Assume that the representation $T(\cdot)$ above is irreducible, and that there is a measure on $G$ which is invariant under left and right shifts. Assume also that convergence conditions are satisfied. Then there exists a measure $\mu$ in $X$ such that we have a resolution of unity

\begin{equation}
\int d\mu (x)|x\rangle\langle x|=I.
\label{u2}
\end{equation}

(See equation (2.3.5) in [11].) The argument behind (\ref{u2}) is essentially the same that I used in [4] to prove the resolution of the identity (where I only assumed left invariance of the measure on $G$): Define an operator $B=\int dx |x\rangle\langle x|$ and show by using Schur's lemma that $B$ must be a constant times the identity operator $I$.

Recall now two crucial assumptions behind the construction above:

1) $\{T(g)\}$ can be chosen as an irreducible representation on some Hilbert space $\mathcal{H}$ of the abstract group behind the group actions $G$. 

2) There is a one-to-one correspondence between the elements $g\in G$, the variable values $\theta$ and the coherent states $T(g)|\theta_0\rangle$.

If $G$ is abelian, it only possible to satisfy 1) if $\mathcal{H}$ is one-dimensional, giving a trivial theory.

But assuming 1) and 2), one can now, again under convergence conditions, define a unique operator in $\mathcal{H}$ corresponding to the conceptual variable $\theta$:

\begin{equation}
A^\theta = \int  \theta(x) |x\rangle\langle x | d\mu(x).
\label{u3}
\end{equation}

Properties of such operators are as studied in [4].

\subsection{Two maximal conceptual variables}

For many concrete examples, it is not possible to find a non-abelian group satisfying 1) and 2) above, and then the theory of the previous subsection is just trivial. For this case, study \emph{two} conceptual varables $\theta$ and $\xi$. 

Assume that the variables $\theta$ and $\xi$ are maximally accessible, that both can be seen as functions of an underlying inaccessible variable $\phi$, and suppose that there exists a transformation $k$ such that $\xi (\phi)=\theta (k\phi)$. Let $g\in G$ be a transitive  group action on $\theta$, and let $h\in H$ be the transitive group action on $\xi$  
defined by $h\xi (\phi)=g^1 \theta (k\phi)$ when $\xi(\phi)=\theta(k\phi)$, where $g^1\in G^1$, an independent copy of $G$. This gives a group isomorphism between $G$ and $H$.

Let $n\in N$ be the group actions on $\psi =(\theta ,\xi )$ generated by $G$ and $H$ and a single element $j$ defined by $j\psi =(\xi,\theta)$ and $j\theta =\xi$. For $g\in G$, define $g j\psi (\phi) =(g\theta (k\phi),\theta(\phi))$ when $\xi =\theta (k\phi)$, and for $h\in H$ define $h j\psi =(\xi (\phi), h \xi (k^{-1}\phi))$ when $\theta(\phi)=\xi (k^{-1}\phi)$. Since $G$ and $H$ are transitive on the components, and since through $j$ one can choose for a group element of $N$ to act first arbitrarily on the first component and then arbitrarily on the second component, $N$ is transitive on $\psi$. Also, $N$ is non-Abelian: $gj\ne jg$.

I  want to fix some Hilbert space $\mathcal{H}$, and define a representation $U(\cdot)$ of the group corresponding to $G$ on this Hilbert space  with the property that if we fix some vector $|v_0\rangle\in\mathcal{H}$, then the vectors $U(g)|v_0\rangle$ are in one-to-one correspondence with the group elements $g\in G$ and hence with the values $g\theta_0$ of $\theta$ for some fixed $\theta_0$. 

For each element $g\in G$ there is an element $h =j g j\in H$ and vise versa. Note that $j\cdot j=e$, the unit element. Let $U(j)=J$ be some unitary operator on $\mathcal{H}$ such that $J\cdot J =I$. Then for the representation $U (\cdot)$ of the group corresponding to $G$, there is a representation $V (\cdot)$ of the group corresponding to $H$ given by $V (j gj)=J U (g)J$. These are acting on the same Hilbert space $\mathcal{H}$ with vectors $|v\rangle$, and they are equivalent in some concrete sense.

Note that $J$ must satisfy  $J U(j g j)=U (g)J$.  By Schur's Lemma this demands $J$ to be an isomorphism or the zero operator if the representation $U(\cdot )$ is irreducible. In the reducible case a non-trivial operator $J$ exists, however:

In this case there exists at least one proper invariant subrepresentation $U_0$ acting on some vector space $\mathcal{H}_0$, a subspace of $\mathcal{H}$, and another proper invariant subrepresentation $U'_0$ acting on an orthogonal vector space $\mathcal{H}'_0$. Fix $|v_0\rangle \in \mathcal{H}_0$ and $|v'_0\rangle \in \mathcal{H}'_0$, and then define $J|v_0\rangle=|v'_0\rangle$, $J|v'_0\rangle=|v_0\rangle$ and $J|v\rangle=|v\rangle$ for any $|v\rangle\in \mathcal{H}$ which is orthogonal to $|v_0 \rangle$ and $|v'_0 \rangle$.

Now we can define a representation $W(\cdot)$ of the full group $N$ acting on $\psi =(\theta, \xi)$ in the natural way: $W(g)=U(g)$ for $g\in G$, $W(h)=V(h)$ for $h\in H$, $W(j)=J$, and then on products from this.

If $U$ is irreducible, then also $V$ is an irreducible representation of $H$, and we can define operators $A^\theta$ corresponding to $\theta$ and $A^\xi$ corresponding to $\xi$ as in (\ref{u3}). If not, we need to show that the representation $W$ of $N$ constructed above is irreducible on $\mathcal{H}$.
\bigskip

\textbf{Lemma 1.} \textit{$W(\cdot)$ as defined above is irreducible.}
\bigskip

\textit{Proof.}
Assume that $W(\cdot)$ is reducible, which implies that both $U(\cdot)$ and $V(\cdot)$ are reducible, i.e., can be defined on a lowerdimensional space $\mathcal{H}_0$, and that $E=W(j)$ also can be defined on this lower-dimensional space. Let $R(\cdot)$ be the representation of $G$ restricted to vectors $|u\rangle$ in $\mathcal{H}$ orthogonal to $\mathcal{H}_0$. Fix some vector $|u_0\rangle$ in this orthogonal space; then consider the vectors in this space given by $R(g)|u_0\rangle$. Note that the vectors orthogonal to $\mathcal{H}_0$ together with the vectors in $\mathcal{H}_0$ span $\mathcal{H}$, and the vectors $U(g)|u_0\rangle$ in $\mathcal{H}$ are in one-to-one correspondence with $\theta$. Then the vectors $R(g)|u_0\rangle$. are in one-to-one correspondence with a subvariable $\theta^1$. And define the representation $S(\cdot)$ of $H$ by $S(jgj) =R(g)$ and vectors $S(h)|v_0\rangle$, where $|v_0\rangle$ is a fixed vector of $\mathcal{H}$, orthogonal to $\mathcal{H}_0$. These are in one-to-one correspondence with a subparameter $\xi^1$ of $\xi$.

Given a value $\theta$, there is a unique element $g_\theta \in G$ such that $\theta = g_\theta \theta_0$. (It is assumed that the isotropy group of $G$ is trivial.)

From this look at the fixed vector $S(jg_\theta j)|v_0\rangle$. By what has been said above, this corresponds to a unique value $\xi^1$, which is determined by $g_\theta$, and hence by $\theta$. But this means that a specification of $\theta$ determines the vector $(\theta, \xi^1)$, contrary to the assumption that $\theta$ is maximally accessible. Thus $W(\cdot)$ cannot be reducible.

\qed

This shows that there are group actions $n\in N$ acting on $(\theta,\xi)$ and an irreducible representative $W$ of the corresponding group acting on some Hilbert space $\mathcal{H}$. Hence the conclusion (\ref{u2}) holds if $G$ above is replaced by $N$ and the irreducible representation is $W(\cdot)$. That is, the crucial assumption 1) is now satisfied. It is left to prove that the Hilbert space $\mathcal{H}$ can be chosen so that condition 2) holds in this case.

Now $N$ is generated by $G$, $H$ and a group $L$ with two elements $l$, the identity element and $j$. Define a binary varable $\lambda$ such that $\lambda=0$ if $l$ is the identity, $\lambda=1$ if $l=j$. First I will establish a one-to-one correspondence between the values of $(\theta, \xi ,\lambda)$ and the elements $n$ of the group $N$. Now these elements are partly constructed from group elements $g$ acting on $\theta$, and since $G$ is assumed to be transitive with a trivial isotropy group, there is a one-to-one correspondence between $g$ and $\theta$. The group $H$ constructed in the proof is also transitive on $\Omega_\xi$ with a trivial isotropy group, so there is a one-to-one correspondence between the values $\xi$ and the group elements $h\in H$. Finally, there is a one-to-one correspondence between $\lambda$ and $l$. But $n\in N$ as acting on $(\theta,\xi, \lambda )$ is given as $n=(g,h,j)$, so the required one-to-one correspondence is established.

It is only left to prove that under suitable assumptions the vectors $W(n)|\nu_0\rangle$ are in one-to-one correspondence with the group elements $n$, and hence with $\zeta=(\theta,\xi,\lambda )$. Once this is proved, by the construction of subsection A  one can define coherent states $|\psi\rangle =W(n)|\psi_0\rangle\in \mathcal{H}$ and a left invariant  measure $v(\psi)$ such that

\begin{equation}
\int |\psi\rangle\langle\psi |\nu (d\psi)=I.
\label{u4}
\end{equation}

As above, with a change of notation, $\psi$ are elements of the homogeneous space $\Psi = N/M$, where $M$ is the isotropy subgroup of the group $W(N)$ corresponding to the initial state $|\psi_0\rangle$. In other words it is the maximal sugroup $M=\{m\}$ such that

\begin{equation}
W(m)|\psi_0\rangle=\mathrm{exp}[\alpha(m)]|\psi_0\rangle.
\label{u5}
\end{equation}

Now $N$ is a group acting on $(\theta,\xi,\lambda )$ constructed from the groups $G$ acting on $\theta$ and $H$ acting on $\xi$ and an element $j$ such that $j(\theta,\xi)=(\xi, \theta)$.

We want to characterize the elements $z$ of $N/M$. As in (\ref{u1}) the elements of $M$ are such that (\ref{u5}) holds.
\bigskip

\textbf{Lemma 2.} \textit{We can write $z =(x,y,l)$, where $x$ is an element of the homogeneous space $X=G/E$,  $y$ is an element of the homogeneous space $Y=H/F$, and $l$ takes the values 0 and 1. Here $E$ is the isotropy subgroup of the group $U(G)$, and $F$ is the isotropy subgroup of the group $V(H)$, both corresponding to the initial state $|\psi_0\rangle\in\mathcal{H}$.}
\bigskip

\textit{Proof.}
For $e\in E$ we have

\begin{equation}
W(e)|\psi_0\rangle =\mathrm{exp}[i\alpha(e)]|\psi_0\rangle,
\label{u01}
\end{equation}

while for $f\in F$ we have

\begin{equation}
W(f)|\psi_0\rangle =\mathrm{exp}[i\alpha(f)]|\psi_0\rangle.
\label{u02}
\end{equation}

Normalizing $\alpha(m)$ for each $m$ such that $\alpha(j)=0$, (\ref{u5}) follows from (\ref{u01}) and (\ref{u02}).

Now the cosets $nM$ are found by first specializing $n$ to $g$ and $h$, respectively $j$. We have $j(\theta,\xi)=(\xi,\theta)$, and the cosets $gM$ and $hM$ are given in terms of $x$ and $y$. It follows that $z=(x,y,l)$ if we in addition allow $jM$ to interchange $x$ and $y$.

\qed

\subsection{The operators}

By the symmetry of the situation, and since $G$ and $H$ act independently on $\theta$ and $\xi$ in the construction of the group $N$, we conclude that the measure $\nu$ in (\ref{u4}) can be written as $\nu (d\psi )=\rho (dx)\rho(dy)\omega(l)$ for some marginal measure $\rho$, where $\omega$ is any measure on the two variables 0 and 1. Then we can define

\begin{equation}
P(x)=\sum_l \int_Y  |\psi\rangle\langle\psi| \rho(dy)\omega(l),
\label{u6}
\end{equation}

and

\begin{equation}
A^\theta = \int_X \theta(x) P(x) \rho (dx).
\label{u7}
\end{equation}

Similarly:

\begin{equation}
Q(y)=\sum_l \int_X  |\psi\rangle\langle\psi| \rho(dx)\omega(l),
\label{u8}
\end{equation}

and

\begin{equation}
A^\xi = \int_Y \xi(y) Q(y) \rho (dy).
\label{u9}
\end{equation}

It follows from (\ref{u4}) that $A^\theta=I$ when $\theta$ is identically equal to 1.

This define the operators for every pair of maximally accessible conceptual variables $\theta$ and $\xi$. The operators for variables that are not maximal, are found by using the spectral theorem, taking as a point of departure the operator for a corresponding maximal variable (cp. equations (4.28) and (4.30) in [4]). 

Again it is an open problem to prove the general uniqueness of these operators. For the spin component case, explicit formulas implying the uniqueness are derived in [4], [11] and [13].

\section{Properties of the operators}

To continue the theory, we prove the important result (cp. Theorem 4.2 in [4]):
\bigskip

\textbf{Theorem 2.}
\textit{Assume that the function $\theta(\cdot)$ is permissible with respect to a group $K$.  For any transformation $t\in K$ and any unitary representation $V$ of $K$, the operator $V(t)^\dagger A^\theta V(t)$ is the operator corresponding to $\theta'$ defined by $\theta'(\phi)=\theta(t\phi)$.}

\bigskip

\textit{Proof.}
By (\ref{u7}) we have
\begin{equation}
V(t^{-1})A^\theta V(t)=\int_X \theta(x) P_t(x)\rho (dx),
\label{u10}
\end{equation}
where
\begin{equation}
P_t(x)=\sum_l \int_{Y} |\psi(t^{-1}\zeta)\rangle\langle \psi(t^{-1}\zeta)| \rho (dy)\omega(l).
\label{u11}
\end{equation} 

Here $|\psi(t^{-1}\zeta)\rangle$ are the  ket vector is constructed such that they are in one-to-one correspondence with $t^{-1}\phi=t^{-1}(\theta,\xi)$ and hence with $t^{-1}(g,h)$. Note that $t^{-1}\phi$ and $\theta=\theta(\phi)$, $\xi=\xi(\phi)$ defines $t^{-1}\zeta$. Write the  group elements $t^{-1}(g,h)$ as $(g',h')$, constituting groups $G'$ and $H'$ acting on $\theta$ and $\xi$, respectively. Let $E'$ be the subgroup of $G'$ constructed as in (\ref{u1}), and let $F'$ be the corrsponding subgroup of $H'$. Write $X'=G'/E'$ and $Y'=H'/F'$. The elements of these cosets may be defined as $x'=t^{-1}x$ and $y'=t^{-1}y$, respectively. By the left invariance of the measure $\rho$,  (\ref{u10}) and (\ref{u11}) can be written as

\begin{equation}
V(t^{-1})A^\theta V(t)=\int_X \theta(x) P_t(x)\rho (dt^{-1}x),
\label{u12}
\end{equation}
where
\begin{equation}
P_t(x)=\sum_l \int_{Y} |\psi(t^{-1}\zeta)\rangle\langle \psi(t^{-1}\zeta)| \rho (dt^{-1}y)\omega(l).
\label{u13}
\end{equation} 

The last integral may be written

\begin{equation}
P_t(x)=\sum_l \int_{Y} |\psi(\theta(t^{-1}\phi), \xi(\phi')\rangle\langle \psi(\theta(t^{-1}\phi), \xi(\phi')| \rho (dy')\omega(l),
\label{u14}
\end{equation} 
and we can write $P_t(x)=P(t^{-1}x)$. This is inserted into (\ref{u12}), and using left-invariance of the measure again, this give that the operator $V(t^{-1})A^\theta V(t)$ is associated with the conceptual variable $\theta (tx)$, which also may be written as $\theta(t\phi)$.
\qed

\bigskip

By using this result in the same way as Theorem 4.2 is used in [4], a rich theory follows. I will limit me here to the case where $\theta$ is a discrete conceptual variable. Then one can show:

1) The eigenvalues of $A^\theta$ coincide with the values of $\theta$.

2) The variable $\theta$ is maximally accessible if and only if the eigenvalues of $A^\theta$ are non-degenerate.

3) For the maximal case the following holds in a given context: a) For a fixed $\theta$ each question `What is the value of $\theta$?' together with a sharp answer `$\theta=u$' can be associated with a normalized eigenvector of the corresponding $A^\theta$. b) If there in the context is a set $\{\theta^a ; a\in \mathcal{A}\}$ of maximally accessible conceptual variables (these must by the results of Section IV be related to each other) one can consider all ket vectors that are normalized eigenvectors of some operator $A^{\theta^a}$. Then each  of these may be associated with a unique question-and-answer as above. 

\section{The spin case}

Let $\phi$ be the total spin vector of a particle, and let $\theta^a =\|\phi\| \mathrm{cos}(\phi ,a)$ be the component of $\phi$ in the direction $a$. If we have a coordinate system, the components $\theta^x , \theta^y$ and $\theta^z$ are of special interest. As noted in [4], these components are not permissible if $K$ is the rotation group for some fixed $\|\phi\|$.

Consider a Stern-Gerlach experiment with a beam of particles in the $y$ direction. Write $\phi=(\phi^1 ,\theta^y )$ where $\phi^1$ is the spin component in the $xz$-plane. Let $K_0$ be the group of rotations of $\phi^1$ in this plane for fixed $\|\phi^1\|$.
\bigskip

\textbf{Proposition 2.} \textit{Any component $\theta^a$ in the $xz$-plane is a permissible function of $\phi^1$ under $K_0$.}
\bigskip

\textit{Proof.}
It is sufficient to consider $\theta^x$. Let $k_0\in K_0$ and $\theta^x (\phi^1_1 )=\theta^x (\phi^1_2)$. The group element $k_0$ rotates the $x$-axis to a new direction $x_0$ and $\phi^1_1$ and $\phi^1_2$ to new vectors   $k_0 \phi^1_1$ and $k_0 \phi^1_2$. By the geometry, the angle $(x,\phi^1_1 )$ must be equal to angle $(x_0 , k_0 \phi^1_1 )$, and angle $(x,\phi^1_2 )$ must be equal to angle $(x_0 , k_0 \phi^1_2 )$. Because the cosines of the unrotated angles are equal by assumption, the cosines of the rotated angles must also be equal. $\Box$
\smallskip

Unit vectors in a plane as quantum state vectors have been discussed extensively in [14]. These vectors can be seen as elements of a real 2-dimensional Hilbert space, and they are uniquely determined by the angle they form with the $x$-axis. Using this as a point of departure, density matrices and quantum operators (matrices) were defined in [14]. The authors noted that all real 2-dimensional matrices were generated by the unit matrix and the two real Pauli spin matrices $\sigma_1$ and $\sigma_3$.

Spin coherent states and their connection to the group SU(2) are treated for instance in [11] and [13]. I will follow parts of [5] without going into details. It is crucial that any irreducible representation $D$ is given by a nonnegative integer or half-integer $r$: $D(k)=D^r (k)$, $\mathrm{dim}D^r =2r+1$. In the representation space $\mathcal{H}=\mathcal{H}_r$ the canonical basis $|r;m\rangle$ exists, where $m$ runs from $-r$ to $r$ in unit steps. The infinitesimal operators $A^{\pm}=A^x \pm A^y$, $A^0 =A^z$ of the group representation $V^r$ satisfy the commutation relations

\begin{equation}
[A^0 , A^{\pm}]=\pm A^{\pm},\ \ [A^- ,A^+ ]=-2 A^0 .
\label{commutation}
\end{equation}
The operators  $A^x$, $A^y$ and $A^z$ are related to infinitesimal rotations around the $x$-axis, $y$-axis and $z$-axis, respectively, and may be shown to be identical to the operators associated with $\theta^x$, $\theta^y$ and $\theta^z$, as considered in the previous sections here. We take $\bm{A}=(A^x ,A^y ,A^z )$. The representation space vectors $|r;m\rangle$ are eigenvectors for the operators $A^0$ and $\bm{A}^2 = (A^x)^2 +(A^y)^2 + (A^z)^2$:

\begin{equation}
A^0 |r;m\rangle =m|r; m\rangle ,\ \ \bm{A}^2 |r;m\rangle =r(r+1)|r;m\rangle .
\label{eigen}
\end{equation} 

The operator $\mathrm{exp}[i\omega(\bm{n}\cdot\bm{A})]$, $\|\bm{n}\| =1$, describes the rotation by the angle $\omega$ around the axis directed along $\bm{n}$. In [5] this was used to describe the coherent states $D(k)|\phi_0\rangle$ in various ways. The ket vector $|\phi_0\rangle$ may be taken as $|r;m\rangle$ for a fixed $m$; the simplest choice is $m=-r$. 

The spin case is discussed more thoroughly in [4]. Among other things, the discretization of spin components is motivated from a general principle of model reduction.

\section{The epistemic interpretation}

It is very confusing to people outside the quantum foundation community that quantum mechanics has many different, mutually excluding interpretations. A recent Wikipedia article lists 16 different interpretations. The mathematical foundation developed above and in [4] implies a rather wide interpretation, generalizing the QBist interpretation and related to several of the other interpretations. I will call this the general epistemic interpretation.

Consider a physical system, and an observer  or a set communicating observers on this system. The physical variables which can be measured  in this setting are examples of accessible conceptual variables, and are called e-variables in [4] and [8].

The approach here could be related to parts of various recent derivations of quantum theory from a set of postulates (see, for instance, [15] and [16]). As is stated in [17], there is a problem connecting these derivations to the many different interpretations of quantum theory. By contrast, the derivation presented here is tied to a particular interpretation, namely  the general epistemic interpretation. This is also elaborated on in [4].

A maximally accessible variable $\theta^a$ admits values $u_j^a$ that are single eigenvalues of the operator $A^a$, uniquely determined from $\theta^a$. Let $|a;j\rangle$ be the eigenvector associated with this eigenvalue. Then $|a;j\rangle$ can be connected to the question `What is the value of $\theta^a$?' together with the sharp answer `$\theta^a =u_j^a$'. 

A general ket vector $|v\rangle\in\mathcal{H}$ is always an eigenvector of \emph{some} operator associated with a conceptual variable. It is natural to conjecture that this operator at least in some cases can be selected in such a way that the accessible variable is maximally accessible. (For the general problem, see [4].) Then $|v\rangle$ is in a natural way associated with a question-and-answer pair. It is of  interest that H\"{o}hn and Wever [18] recently derived quantum theory for sets of qubits from such question-and-answer pairs; compare also the present derivation. Note that the interpretation implied by such derivations is relevant for both the preparation phase and  the measurement phase of a physical system.

One may also consider linear combinations of ket vectors from this point of view. Let the maximally accessible variable $\theta^b$ admits values $u_i^b$ that are single eigenvalues of the operator $A^b$, uniquely determined from $\theta^b$. Let $|b;i\rangle$ be the eigenvector associated with this eigenvalue. Then one can write
\begin{equation}
|b;i\rangle=(\sum_j |a;j\rangle\langle a;j|)|b; i\rangle =\sum_j \langle a;j|b;i\rangle |a;j\rangle .
\label{u0}
\end{equation}

This state can be interpreted in terms of a question `What is the value of $\theta^b$?' with the sharp answer `$\theta^b =u_i^b$'. But if one tries to ask questions about $\theta^a$ for a system where the observer or the set of observers is in this state, the answer is simply `I (we) do not know'.

Also, entangled states may be considered from this point of view. Consider two qubits 1 and 2, each having possible spin values in some fixed direction given by +1 and -1, and look at the entangled state
\begin{equation}
|\psi\rangle =\frac{1}{\sqrt{2}}(|1+\rangle\otimes |2-\rangle - |1-\rangle \otimes |2+\rangle).
\label{ent}
\end{equation}

As discussed in [4] and references given there, this is an eigenvector of the operator for the conceptual variable $\delta =\theta_x\eta_x+\theta_y\eta_y+\theta_z\eta_z$ corresponding to the eigenvalue -3. Here $\mathbf{\theta}=(\theta_x ,\theta_y , \theta_z )$ and $\mathbf{\eta}=(\eta_x ,\eta_y , \eta_z )$ are the total spin vectors of the two qubits, and $\delta$ is accessible to an observer observing both qubits. Note that $\delta =-3$ implies $\theta_x\eta_x=\theta_y\eta_y=\theta_z\eta_z=-1$, which again implies that $\theta_a =-\eta_a$ for any direction $a$. It is trivial that this state can be interpreted in terms of the question `What is $\delta$?' together with the sharp answer `$\delta=-3$'. (The other possible value for $\delta$ is -1, which corresponds to a three-dimensional eigenspace for the operator.)

From a general point of view it may be considered of some value to have an epistemic interpretation which is not necessarily tied to a strict Bayesian view (see for instance [19]  on this). Under an epistemic interpretation, one may also discuss various ``quantum paradoxes'' like Schr\"{o}dinger's cat, Wigner's friend and the two-slit experiment. 

In Example 1 and Example 2 below I limit myself to quantum states that can be interpreted in terms of question-and-answers for maximally accessible variables. In the contexts discussed, this gives an understanding where the relevant observers are not able to distinguish between superposition and mixture.
\bigskip

\textbf{Example 1.} \textit{Schr\"{o}dinger's cat.} The discussion of this example concerns the state of the cat just
before the sealed box is opened. Is it half dead and half alive?

To an observer outside the box the answer is simply: ``I do not know''. Any accessible e-variable
connected to this observer does not contain any information about the status of life of the cat. But on
the other hand – an imagined observer inside the box, wearing a gas mask, will of course know the
answer. The interpretation of quantum mechanics is epistemic, not ontological, and it is connected to the
observer. Both observers agree on the death status of the cat once the box is opened.
\bigskip

\textbf{Example 2.} \textit{Wigner’s friend.} Was the state of the system only determined when Wigner learned the
result of the experiment, or was it determined at some previous point?

My answer to this is that at each point in time a quantum state is connected to Wigner’s
friend as an observer and another to Wigner, depending on the knowledge that
they have at that time. The superposition given by formal quantum mechanics corresponds to a `do not
know' epistemic state. The states of the two observers agree once Wigner learns the result of the
experiment.
\bigskip

\textbf{Example 3.} \textit{The two-slit experiment.} This is an experiment where all real and imagined observers
communicate at each point of time, so there is always an objective state. 

Look first at the situation when we do not know which slit the particle goes through. This is 
a `do not know' situation. Any statement to the effect that the particles somehow pass through both
slits is meaningless. The interference pattern can be explained by the fact that the particles are (nearly)
in an eigenstate in the component of momentum in the direction perpendicular to the slits in the plane
of the slits. If an observer finds out which slit the particles goes through, the state changes into an
eigenstate for position in that direction. In either case the state is an epistemic state for each of the
communicating observers, which might indicate that it in some sense can be seen as an ontological state. But this must be seen as a state of the screen and/or the device to observe the particle, not as an ontological state of the particle itself.

\section{Discussion}

Some of the results above were limited to accessible conceptual variables taking a finite number of values. But the notion of conceptual variables is much more general, and can thus be considered in more general settings.

This notion of conceptual variables also has links to other interpretations of quantum theory. Take for instance the classical Bohm interpretation, constructed from particle trajectories plus a pilot wave. These constructions may be seen as conceptual variables, but at least the full trajectory must be inaccessible. Or consider the many worlds interpretation, where the variables associated with the different worlds must be considered as conceptual variables, but only one world is accessible at each point in time.

The arguments developed here are based on focusing (choice of maximally accessible conceptual variable) and symmetry. These concepts are not confined to the microscopic world. This is consistent with the fact that quantum theory has recently been applied to cognitive models and to certain social and economic models (see [6] and [20] and references there). In fact it has been claimed that the quantum structure is ubiquitous, see [21].

Further philosophical consequences of these results are discussed in Chapter 6 of [4]. In general it is referred to [4] for further discussions.

It is quite crucial that the proof of Theorem 1 as given here is correct. This theorem not only serves as a basis for the derivation of quantum theory as discussed in [4], it has also now  been used in a discussion of the Bell theorem, the Bell experiment and a general assertion of our limitation as human beings when making decisions [10]. I am currently working with using the same theorem as a point of departure for discussing when a given quantum state also has an ontologic interpretation and the more speculative question of whether \emph{all} quantum states can be said in some sense to have ontological interpretations. Note that in [4] and in the present paper, the initial interpretation of states is epistemological.

One important background for this whole development has been that, in my opinion there was too little communication between researchers working with the foundation of quantum theory and researchers in the statistics society. My book has been an attempt to develop elements of a future common culture.

What is culture? According to the author and philosopher Ralph D. Stacey it is a set of attitudes, opinions and convictions that a group of people share, about how one should act towards each other, how things should be evaluated and done, which questions that are important and answers that may be accepted. The most important elements in a culture are unconscious, and cannot be forced upon one from the outside.

One hope now, is that versions of results like Theorem 1 above, will be background results which can function as a part of a common scientific culture covering both quantum physicists and statisticians. If this happens, I feel that it then also should be easier to arrive at some joint understanding of physical theories describing the microscopic world and theories describing the macroscopic world.

In my opinion it is very important in the long run to try to develop a common international cultural foundation for all scientists. In this way, science could really achieve the authority that it requres and needs. There are extremely many problems facing the world now: climate, health problems, poverty problems, refugee problems, international conflicts and the existence of very dangerous weapons. For all these problems one should try to device rational solutions that at the same time satify good ethical standards; in particular, rational and ethical decisions should be made by national and international leaders. Unfortunately, in the processes of making decisions, we may all be limited. (In a concrete quantum situation, this has bee argued for in [10]). So, ideally, to arrive at good decisions, in addition to other requirements, our leaders should try to get  rational inputs. Ideally again, such inputs could in each concrete case be given by joint efforts by groups of scientists.

In improving this situation, we may all be somewhat responsible. In particular in this connection, the process of achieving knowledge is important. And, again in my view, such a process can always be associated with the mind of a single person or by the joint minds of a group of communicating persons. Knowledge as such may or may not be coupled directly to an external objective real world.

In his book [22] Lee Smolin has listed 5 great problems in theoretical physics. Combining the conclusions of [4], [10] and the present article seems to me to be a way to device a solution to at least a large part of problem 2:  Resolve the problems in the foundation of quantum mechanics. This is done partly through appealing to a particular unification: Try to find a  unified new culture that is meaningful both to physicists and to statisticians. As I see it, this may also perhaps be a stepping stone for approaching further problems, like finding connections between elementary quantum theory and other theories, in particular field theory, theories in particle physics and relativity theory. But all this belongs to the future.

In the early book by David Bohm and F. David Peat [23] it was emphasized in a physical research setting that a conceptual discussion ought to precede formal mathematical calculations. I am grateful to Solve S\ae b\o \ for making me aware of that book. It is very interesting, but this does not mean that I agree with David Bohm in everything he says. In particular, my general views of the interpretation of quantum mechanics are more fully explained in [4].

Just as I was about to submit this paper, I received Lee Smolin's last book [24]. Having looked briefly on that book, I want to stress an obvious fact: I do not claim in any way that my approach will solve all problems in connection to quantum mechanics. But I do claim that my Theorem 1 implies a much simpler foundation and interpretation, and that this in itself may facilitate somewhat discussions of many questions. 

Smolin declares himself as a realist, and identifies his position by saying yes to the following two questions. I will partly describe my own position by trying to answer the same questions.

1) Does the natural world exist independently of our minds? More precisely, does matter have a stable set of properties in and of itself, without regards to our perception and knowledge?

2) Can those properties be comprehended and described by us? Can we understand enough of the laws of nature to explain the history of our universe and predict its future?

To the first part of 1) I will answer a clear yes. I think one must be very theoretically minded to doubt the existence of the natural world. But when it comes to properties, I will tend to say no, if we should think of all possible properties in some given physical situation. A critical situation is the Bell experiment, discussed in [10]. I think the answer to this part of the question has to be related to our philosophy.

To take another example, consider an observer who has the choice between measuring the position or the momentum of some particle. I partly adher to the philosophy of Zwirn [25], who claims that all description of the world must be relative to the mind of some person. But persons can communicate. The mind of the observer in this case fits very well into the situation described in Theorem 1 here. Hence it can be coupled to a quantum formulation. He can find a very accurate value for position, similarly for momentum, but not for both. Hence one has a very peculiar situation. The position may `exist' for him; the momentum may `exist' for him, but not both at the same time. Following [25], this is all one can say about existence.

Can the situation be saved by considering a group of communicating observers? Again I will say no. Since they are communicating before the measurement, they must decide on the same measurement, and hence arrive at the same conclusion regarding `existence'.

I disagree slightly with Zwirn when it comes to groups of communicating people. I will claim that since they as above make common decisions on measurement, regarding to the existence question, they can be seen as  one person. In particular, they have agreed on common conceptual variables `position' and `momentum', and can according to Theorem 1 have a common coupling to quantum theory. 

To the first part of 2) I will say yes, and to the second part no, at least to some extent. This again can be referred to [10], where I conclude that we all may be limited when making decisions, in particular the decisions that we may have to make in our description of the properties of the world. Of course, to some degree we can explain the history of our universe and predict its future. Modern physics and astronomy is impressive here. But we can never go in complete details, at least when predictions are concerned.

\section*{References}

[1] Ballentine, L.E. (1998). \textit{Quantum Mechanics. A Modern Development.} World Scientific, Singapore.\newline

[2] Takhtajan, L.A. (2008). \textit{Quantum Mechanics for Mathematicians.} Graduate Studies in Mathematics \textbf{95}, American Mathematical Society.\newline

[3] Hall, B.C. (2013). \textit{Quantum Theory for Mathematicians.} Graduate Course in Mathematics. Springer, New York. \newline

[4] Helland, I.S. (2021). \textit{Epistemic Processes. A Basis for Statistics and Quantum Theory.'} Revised version. Springer, Berlin. \newline

[5] Helland, I,S.(2010)  \textit{Steps Towards a Unified Basis for Scientific Models and Methods.} World Scientific, Singapore.\newline

[6] Busemeyer, J.R. and Buza, P.D. (2012). \textit{Quantum Models for Cognition and Decision.} Cambridge University Press, Cambridge.\newline

[7] Yukalov, V.I. and D. Sornette, D.(2010) Mathematical structure of quantum decision theory. \textit{Adv. Compl. Syst.} \textbf{13}, 659--698.\newline

[8] Helland, I.S. (2019a). Symmetry in a space of conceptual variables. \textit{Journal of Mathematical Physics} \textbf{60} (5) 052101. Erratum \textit{J. Math. Phys.} \textbf{61} (1) 019901 (2020).\newline

[9] Helland, I.S. (2019b). When is a set of questions to nature together with sharp answers to those questions in one-to-one correspondence with a set of quantum states? arXiv: 1909.08834 [quant-ph].\newline

[10] Helland, I.S. (2021). The Bell experiment and the limitation of actors. To appear in Foundations of Physics.  \newline

[11] Perelomov, A. (1986). \textit{Generalized Coherent States and Their Applications.} Springer, Berlin.\newline

[12] Vaidman, L. (2019). Derivations of the Born rule.  phil-sci-archive.pitt.edu (2019).\newline

[13] Gazeau, J.-P. (2009). \textit{Coherent States in Quantum Physics.} Wiley-VCH, Weinberg.\newline

[14] Bergeron, H., Curado, E.M.F.,  Gazeau, J.-P.  and L. M. C. S. Rodrigues, M.C.S. (2017). A baby Majorana quantum formalism.'' arXiv: 1701.04026v3 [quant-ph]. \newline

15] Hardy, L. (2001). Quantum theory from reasonable axioms. arXiv: 01010112v4 [quant-ph] (2001).\newline

[16] Chiribella, G., D'Ariano, G.M, and Perinotti, P. (2016).  Quantum from principles. In: \textit{Quantum Theory: Informational Foundation and Foils.} Chiribella, G. and Spekkens, P.W. [Eds.] pp. 171-221. Springer, Berlin. \newline

[17] Chiribella, G.,  Cabello, A.,  Kleinmann, M.  and M\"{u}ller, M.P. (2019).  General Bayesian theories and the emergence of the exclusitivity principle. arXiv: 1901.11412v2 [quant-ph].\newline

[18] H\"{o}hn, P.A. and Wever, C.S.P. (2017). Quantum theory from questions.  \textit{Physical Review} A \textbf{95}, 012102 (2017)\newline

[19] Fuchs, C.A. (2010). QBism, the perimeter of quantum Bayesianism. arXiv: 1003.5209 [quant-ph]. \newline

[20] Haven, E. and Khrennikov, A. (2013). \textit{Quantum Social Sciences.} Cambridge Press, Cambridge.\newline

[21] Khrennikov, A. (2010). \textit{Ubiquitous quantum structure: from psychology to finances.} Springer, Berlin.\newline

[22] Smolin, L. (2007). \textit{The Trouble with Physics.} Houghton Mifflin, Boston. (2007). \newline

[23] Bohm, D. and Peat, F.D. (1987). \textit{ Science, Order, and Creativity.} Routledge, London. \newline

[24] Smolin, L. (2019). \textit{Einstein's Unfinished Revolution. The Search for What Lies Beyond the Quantum.} Penguin Books. \newline

[25] Zwirn, H. (2016). The measurement problem: Decoherence and Convivial Solipsism. \textit{Found. Phys.}  \textbf{46}, 635-667.

\section*{Declaration}

There is no funding and no competing interest.

\end{document}